\documentclass[aps,prl,twocolumn,showpacs,preprintnumbers,amsmath,amssymb,superscriptaddress,bibnotes]{revtex4}

\usepackage{natbib}
\usepackage{graphicx}
\usepackage{dcolumn}
\usepackage{bm}
\usepackage{color}

\begin{document}


\title{Strong Coupling Superconductivity in Iron-Chalcogenide  FeTe$_{0.55}$Se$_{0.45}$}

\author{W. K. Park}\email[Electronic address: ]{wkpark@illinois.edu}
\author{C. R. Hunt}
\author{H. Z. Arham}\affiliation{Department of Physics and the Frederick Seitz Material Research Laboratory, University of Illinois at Urbana-Champaign, Urbana, Illinois 61801, USA}
\author{Z. J. Xu}
\author{J. S. Wen}
\author{Z. W. Lin}
\author{Q. Li}
\author{G. D. Gu}\affiliation{Condensed Matter Physics and Materials Science Department,
Brookhaven National Laboratory, Upton, New York, 11973, USA}
\author{L. H. Greene}\affiliation{Department of Physics and the Frederick Seitz Material Research Laboratory, University of Illinois at Urbana-Champaign, Urbana, Illinois 61801, USA}

\date{\today}
             
\begin{abstract}
The superconducting order parameter in the iron-chalcogenide superconductor FeTe$_{0.55}$Se$_{0.45}$ ($T_\textrm{c}$ = 14.2 K) is investigated by point-contact Andreev reflection spectroscopy. The energy gap magnitude (3.8 meV at 1.70 K) and temperature dependence as extracted from the Andreev conductance spectra reveal strong-coupling superconductivity and is consistent with $s$-wave order parameter symmetry. No clear evidence for multiple order parameters or interference from multiple bands is observed. A conductance enhancement persists above $T_\textrm{c}$ to $\sim 18-20$ K and possible origins, including novel quasiparticle scattering due to strong antiferromagnetic spin fluctuations, are discussed.  
\end{abstract}

                            
\maketitle

The recent discovery of superconductivity in Fe-based materials has stimulated research interest in alternate routes to novel superconductivity~\cite{kamihara08}, providing unexpected opportunities for the study of novel and/or unconventional pairing mechanisms. Both similarities to and differences from cuprates and heavy fermions have been unveiled~\cite{mazin09mazin10}. The occurrence of superconductivity in close proximity to antiferromagnetism is reminiscent of their phase diagrams. The antiferromagnetic parent compounds are metallic, unlike the cuprates which are Mott insulators. Also, in Fe-based materials, multiple bands are active in both magnetism and superconductivity, unlike in cuprates. In some sense, these novel materials may act as a bridge between cuprates and heavy fermions~\cite{scalapino10}. Reasonable agreements between electronic structure calculations~\cite{singh08subedi08} and measurements~\cite{xia09} have been observed. The Fermi surface consists of multiple sheets: hole pockets around the (unfolded) Brillouin zone center and electron pockets near the edges. Recent developments have revealed that several degrees of freedom are intertwined to drive magnetic, structural, and pairing instabilities in a complicated manner leading to dependencies on fine-tuned parameters. The interband pairing  mediated by strong spin-fluctuations has been most widely considered as a possible pairing mechanism~\cite{mazin08}. However, two key issues regarding the proposed $s_{\pm}$ superconducting order parameter remain to be resolved: i) the sign reversal between the electron and hole Fermi pockets; and ii) the number of order parameter components.
  
We employ Andreev reflection (AR) spectroscopy to investigate the superconducting order parameters in Fe-based materials (see ref.~\onlinecite{lu10} and references therein). This technique relies on spectroscopic measurements of Andreev conductance across a metallic junction~\cite{park08park09}. Several theoretical calculations have been reported on the conductance characteristics of a junction involving an $s_{\pm}$ symmetry superconductor~\cite{mbtk}. We report on the superconducting energy gap ($\Delta$) in the iron-chalcogenide superconductor FeTe$_{0.55}$Se$_{0.45}$. We find a single $s$-wave order parameter in the strong coupling limit. Clear evidence for multiple gaps or interference effects from multiple bands is not obtained. 

Differential conductance, $G(V)$ $\equiv$ $dI/dV$, across a nanoscale metallic junction formed in our home-built point-contact differential micrometer rig is measured using a standard four-probe lock-in technique. High quality FeTe$_{0.55}$Se$_{0.45}$ single crystals, grown by a horizontal unidirectional solidification method, exhibit resistive superconducting transition onset at 15.6 K and zero-resistance transition at 14.2 K. The latter is taken as $T_\textrm{c}$ here. Bulk superconductivity is confirmed by magnetic property measurements using a SQUID magnetometer. Freshly cleaved (001)-oriented surfaces are used for measurements. Point-contact junctions are formed at low temperature ($< 2$ K) by bringing an electrochemically polished gold tip into gentle contact with the sample. $G(V)$ data are taken  as a function of temperature and magnetic field. Several junctions are tested in each run by moving the tip to different spots without exposing to air. Three different crystal pieces are measured ensuring reproducibility of the features reported here.

\begin{figure}[bp]
\begin{center}
\includegraphics[scale=0.5]{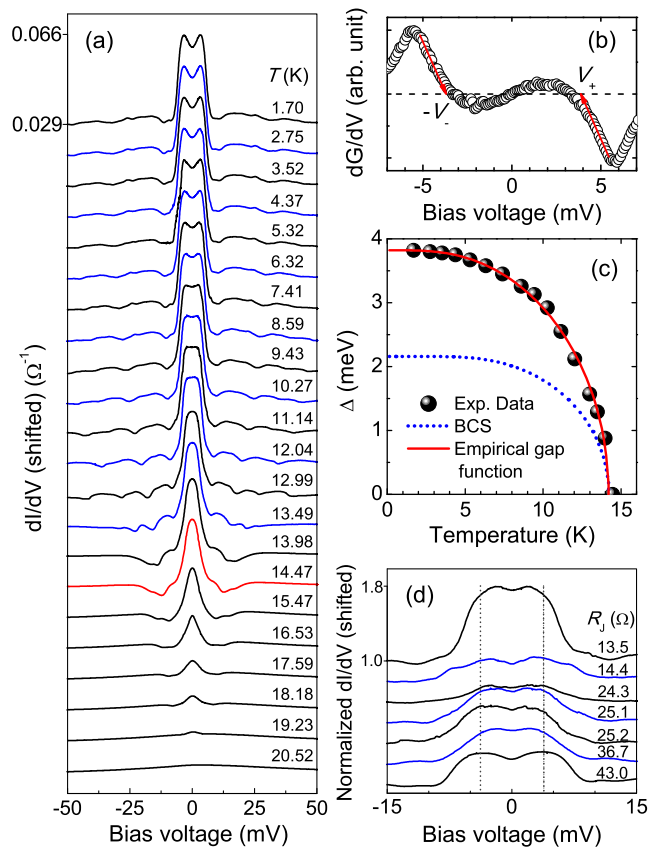}
\end{center}
\vspace{-5pt} \caption{\label{fig1} (color online). (a) Temperature evolution of the conductance spectra of a point-contact junction along the $c$-axis of a FeTe$_{0.55}$Se$_{0.45}$ single crystal. $T_\textrm{c}$ is 14.2 K. (b) $dG/dV$ at 1.70 K to show how $\Delta$ is determined (see text). (c) Extracted $\Delta$ vs. temperature. (d) Conductance spectra at $\sim$ 2 K of several junctions with different resistance ($R_\textrm{J}$), showing similar gap values. In (a) and (d), the indicated $y$-axis scales are for the topmost curves, with other curves shifted downward for clarity by (a) 0.013 $\Omega^{-1}$ and (c) 0.2.}
\end{figure}

Figure~\ref{fig1}(a) shows $G(V)$ spectra of a point-contact junction measured across a bias range of $\pm$ 50 mV from temperature above $T_\textrm{c}$ to 1.7 K. The junction resistance at high bias, $R_\textrm{J}$, varies very little ($<6\%$) over the whole temperature range, indicating the junction remains stable. At low temperature, a conductance enhancement with a double-peak structure is observed around zero-bias due to AR, as predicted by the Blonder-Tinkham-Klapwijk (BTK) model~\cite{blonder82}. With increasing temperature, a single central peak emerges due to thermal smearing. If due solely to superconductivity, this peak would disappear above the bulk $T_\textrm{c}$. As shown in Figs.~\ref{fig1} \& \ref{fig3}, we observe it persists well into the normal state. This robust and reproducible exotic behavior will be discussed later in detail. We first focus on the spectra below $T_\textrm{c}$.

The normalized zero-bias conductance (ZBC) is larger than 2 in the low temperature region, in contradiction with the expectation for the Andreev limit (small barrier strength)~\cite{blonder82}. If we divide out the $G(V)$ data by the one taken just above $T_\textrm{c}$, as typically adopted for normalization in the literature, the resulting conductance peak values are smaller than 2. Thus, we attribute the large ZBC at low temperature to an additional conductance enhancement occurring in the normal state, whose origin is discussed later. Because of this complexity, we leave a complete analysis to future work and focus here on determining $\Delta$ values using a simple scheme: for each curve we take a numerical derivative of the conductance, $dG(V)/dV$, to obtain its two $V$-axis intercepts ($V_{+},-V_{-}$) as demonstrated in Fig~\ref{fig1}(b) for $T=1.70$ K. $e(V_{+} + V_{-})/2$ is taken as $\Delta$. From our experience with standard BTK analysis, we observe that in the Andreev limit smearing factors cause the peaks to move lower than the actual $\pm \Delta / e$. So, the above scheme should provide reasonable and consistent $\Delta$ values, albeit not completely accurate.

The resultant energy gap is plotted in Fig.~\ref{fig1}(c) as function of temperature. A substantial discrepancy is clearly seen between our data and the theoretical prediction for an isotropic s-wave gap in the BCS weak-coupling limit. Instead, the data can be approximated by an empirical gap formula: $\Delta = \Delta_0 \textrm{tanh}[\alpha \sqrt{T_\textrm{c} / T - 1}]$, with $\alpha = 1.55$ (cf. $\alpha = 1.74$ for weak-coupling BCS gap). At the lowest temperature of 1.70 K, our extracted $\Delta$ is 3.80 meV. Then, 2$\Delta / k_\textrm{B} T_\textrm{c} = 6.2 \gg 3.53$, indicating strong coupling superconductivity. Similar $\Delta$ values are reproducibly observed among different junctions on three different crystal pieces, as shown in Fig.~\ref{fig1}(d). Note that different $R_\textrm{J}$ values show similar BTK-like double-peak structures with gap edges around $\pm$3.8 mV, as indicated by the two vertical dashed lines.

Our observation of a single gap in the strong coupling limit is consistent with angle-resolved photoemission~\cite{nakayama09} and neutron scattering~\cite{qiu09} measurements on Fe(Te,Se) crystals with comparable $T_\textrm{c}$s. Much smaller $\Delta$ values are reported in recent scanning tunneling spectroscopy studies~\cite{kato09,hanaguri10}. Holmes {\it et al.}~\cite{homes10} recently claimed the existence of two gaps (2.5 \& 5.1 meV) from an analysis of optical conductivity data on FeTe$_{0.55}$Se$_{0.45}$ crystals (from the same source as ours), in contrast to our observation. We point out that AR spectroscopy is a high energy-resolution technique responding directly and sensitively to the pair potential~\cite{park08park09}. 

While the BTK-like double peak structure is a clear indication of AR-dominant charge transport sensitive to $\Delta$, additional structures with multiple humps and dips are frequently observed outside the gap as seen in Figs.~\ref{fig1}(a) \& (d). We don't take them as a signature for multiple gaps since they are not reproducible from junction to junction. Moreover, we observe (not shown) that those features in the same junction as in Fig.~\ref{fig1} do not show the same magnetic field dependence as the gap: at $T$=1.7 K an applied field of 2 T decreases $\Delta$ from 3.76 meV to 3.54 meV, whereas no measurable change is seen in those additional structures. At present, it is not clear whether they arise from the interference effect predicted by the interfering-band BTK model~\cite{mbtk}.

We now discuss the observability of multiple gaps in multiband superconductors. As is well known, MgB$_2$ is a prototypical superconductor clearly exhibiting two gaps in a variety of measurements. This can be understood as result of fulfilling requirements of both weak-to-no interband pairing interaction and weak interband scattering~\cite{mazin02}. This happens somewhat accidentally in MgB$_2$, where two Cooper pair condensates form on separate parts of the Fermi surface due to the dominant intraband electron-phonon coupling, with simultaneously minimized interband scattering of Cooper pairs. Indeed, electronic structure calculations show that the two bands originating from the hybridization of the boron $p$ orbitals are highly disparate~\cite{choi02}. AR spectroscopy played a central role to establish the multiple superconducting order parameters in MgB$_2$~\cite{szabo01gonnelli02}. For iron-based compounds, multiple bands are involved in magnetism and superconductivity. Learning from the case of MgB$_2$, we can say that whether multiple gaps are observable or not really depends on the nature of pairing interaction and the strength of interband scattering. It is widely considered that a strong interband pairing interaction, which is presumably mediated by spin fluctuations~\cite{mazin08,scalapino10}, is crucial to the unusually high $T_\textrm{c}$s of these materials. According to theoretical calculations for the case of interband pairing, the two gaps tend to merge in the strong coupling limit~\cite{dolgov09stanev08}. Thus, our experimental finding of a single gap in the strong coupling limit can be understood within such theoretical framework. Distinct multiple gaps, if existing at all, should be detectable even in our $c$-axis junctions because of the wide-angle momentum distribution in these metallic junction configurations.

\begin{figure}[bp]
\includegraphics[scale=0.42]{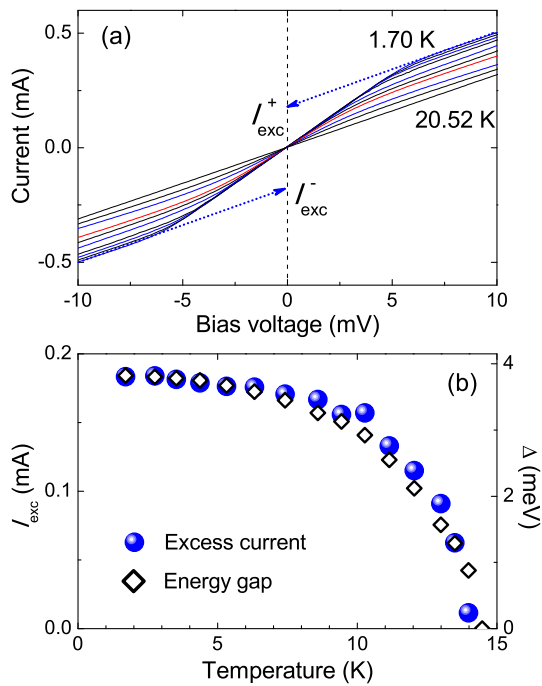}
\vspace{-0pt} \caption{\label{fig2} (color online). (a) $I-V$ characteristics for the same  junction as in Fig.~\ref{fig1}. For clarity, only selected curves are shown: 1.70, 7.31, 9.43, 11.14, 12.99, 13.98, 14.47 (in red), 15.47, 16.53, and 20.52 K from top to bottom. $I_\textrm{exc}^{\pm}$ are excess currents (see text). (b) Excess current, after subtraction of $I_\textrm{exc}$(14.47K), closely follows $\Delta$.}
\end{figure}  

Analysis of $I-V$ characteristics can provide additional information on the superconducting order parameter~\cite{blonder82}. Figure~\ref{fig2}(a) displays $I-V$ curves for the same junction as in Fig.~\ref{fig1}(a). As clearly seen in $G(V)$, a nonlinearity develops around zero-bias as the temperature decreases, which is due to AR. Thus, a quantity, defined as $I_\textrm{exc} \equiv [I_\textrm{NN} - I_\textrm{NS}]_{V \gg \Delta / e}$, is a measure of AR-induced excess current~\cite{blonder82}. In our case, since the $I-V$ curves are slightly asymmetric, we take average of the two excess currents: $I_\textrm{exc} \equiv (I_\textrm{exc}^+ - I_\textrm{exc}^-)/2$, as shown in Fig.~\ref{fig2}(a). The small excess current measured above $T_\textrm{c}$, not due to AR, is subtracted and the resultant $I_\textrm{exc}$ is plotted in Fig.~\ref{fig2}(b). It is clearly seen that $I_\textrm{exc}$ closely follows $\Delta$. This is as expected from the BTK theory for an s-wave superconductor~\cite{blonder82}, showing that the superconducting order parameter in FeTe$_{0.55}$Se$_{0.45}$ is consistent with $s$-wave symmetry.

\begin{figure}[bp]
\includegraphics[scale=0.39]{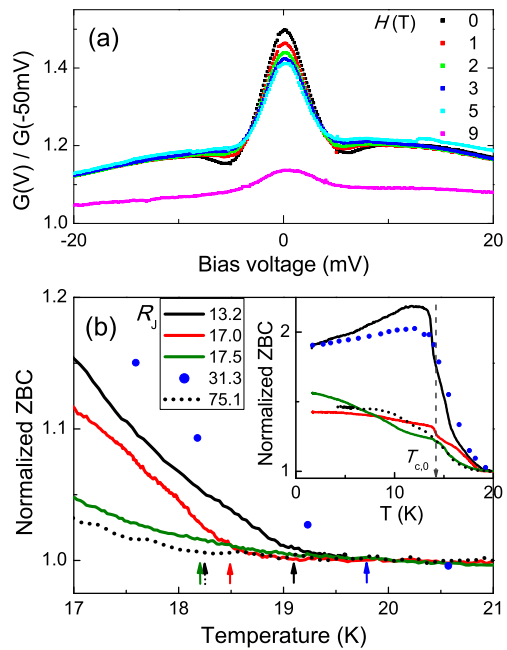}
\vspace{-0pt} \caption{\label{fig3} (color online). (a) Normalized $G(V)$ of a junction at 18.0 K under magnetic field ($H \parallel c$-axis). The abrupt change between 5 and 9 T is due to instability during field ramping. (b) Normalized ZBC of several junctions with different $R_\textrm{J}$(20K). Arrows indicate onset temperatures. The filled circles are from $G(V)$ in Fig.~\ref{fig1}(a) and the dotted line is for a softer Ag paint contact junction. Inset is for an overview.}
\end{figure}

We move on to discuss the conductance features observed in the normal state. As shown in Fig.~\ref{fig1}(a), the ZBC peak persists well above the bulk $T_\textrm{c}$. This behavior is reproducibly observed in all point-contact junctions investigated. Apparently, it doesn't arise from bulk superconductivity since no anomalies are observed in bulk measurements. One may conjecture that the pressure on the sample surface exerted by pressing the tip might enhance the $T_\textrm{c}$ around the junction area. Certainly, this possibility should be taken into consideration because the superconductivity in iron chalcogenides is known to be very sensitive to pressure~\cite{horigane09,gresty09}. 

We have carried out three diagnostic measurements to check this out: i) magnetic field dependence of $G(V)$; ii) dependence of onset temperature on $R_\textrm{J}$; and iii) junctions formed by a softer method. The idea is that if it is due to superconductivity it should disappear above $H_\textrm{c2}$ and show a correlation with $R_\textrm{J}$. Figure~\ref{fig3}(a) shows magnetic field dependence of $G(V)$ at 18.0 K, which is about 10\% lower than the onset temperature for the excess conductance. The $G(V)$ peak is gradually suppressed with increasing field but survives over 9 T. This behavior implies that it may not originate from superconductivity since in the literature $H_\textrm{c2}$ along the $c$-axis of FeTe$_{0.6}$Se$_{0.4}$ crystals at comparable reduced temperature ($T/T_\textrm{c}$) falls to $8 - 8.5$ T range~\cite{yadav09}. To check the relationship between the onset temperature and $R_\textrm{J}$, we compare ZBC curves for several junctions with different $R_\textrm{J}$ values. As shown in Fig.~\ref{fig3}(b), the onset temperature is always in the range of $18 - 20$ K without strong correlation with $R_\textrm{J}$. Note that junctions formed by a softer  Ag paint contact method (dotted line in Fig.~\ref{fig3}(b)), in which the mechanical pressure is expected to be minimal, also exhibit comparable onset temperatures. Combining all these observations, we conclude that the normal state conductance peak is unlikely to be due to superconductivity.

For other possibilities, first we note that it has never been observed in our AR spectroscopy on pure superconducting materials, in which the onset temperature always agrees with the bulk $T_\textrm{c}$. The normal state conductance peak seen in FeTe$_{0.55}$Se$_{0.45}$ is very reminiscent of our previous results on Cd-doped CeCoIn$_5$~\cite{park08physb}, where it persists above bulk $T_\textrm{c}$ (1.3 K) up to $T_\textrm{N}$ (2.9 K), the antiferromagnetic transition temperature. To explain this intriguing behavior, we invoked a novel quasiparticle scattering, as proposed by Bobkova and coworkers~\cite{bobkova05,andersen05}. Here, so-called spin-dependent Q-reflection can occur at an interface with an itinerant antiferromagnet. It is essentially a scattering off an antiferromagnetic order with ordering wave vector, {\bf Q}, just as AR is a scattering off a superconducting pair potential. Detailed calculations~\cite{andersen05} show that the local density of states can exhibit rich structures depending on the junction configuration.

While this scenario appears plausible in systems showing an antiferromagnetic order such as Cd-doped CeCoIn$_5$, it may not be clear how it can be relevant to the Fe(Te,Se) system. Some circumstantial evidence comes from the strong antiferromagnetic spin fluctuations above $T_\textrm{c}$, as observed in nuclear magnetic resonance~\cite{imai09} and neutron scattering~\cite{bao09} measurements. We conjecture that these spin fluctuations in the iron-chalcogenide superconductors might induce a novel quasiparticle scattering process such as the Q-reflection, producing a conductance peak above $T_\textrm{c}$. The kink structures near $T_\textrm{c}$ in the ZBC curves (see the inset of Fig~\ref{fig3}(b)) may be indicative of such scattering actually setting in. While more investigations are necessary to confirm this possibility, we note that our estimated $\Delta$ in Fig.~\ref{fig1} would not be much affected by this (parallel) conduction channel since spin fluctuations are suppressed rapidly below $T_\textrm{c}$~\cite{imai09}.

Finally, we remark on the possibility that the ZBC enhancement is due to the pre-formed Cooper pair state, reported to exist within the pseudogap in cuprates~\cite{hufner08}. To our best knowledge, no AR-like conductance feature has been reported to detect pre-formed pairs. For iron chalcogenides, reported pesudogap-like features seem to be associated with strong spin fluctuations~\cite{imai09,bao09}. Thus, AR from pre-formed Cooper pairs is very unlikely.    

In summary, our AR spectroscopy on FeTe$_{0.55}$Se$_{0.45}$ reveals strong coupling superconductivity. The superconducting energy gap and excess current show temperature dependences for $s$-wave symmetry. No clear evidence for multiple order parameters or interfering bands is observed. A conductance enhancement is reproducibly measured persisting into the normal state to $18 - 20$ K. Our diagnostic measurements rule out pressure-induced local superconductivity. We suggest a novel quasiparticle scattering due to strong spin fluctuations as a strong candidate, whose investigation may help elucidate the pairing mechanism in these novel superconductors.

We thank X. Lu for experimental help, H. Hu and J.-M. Zuo for helpful discussion regarding their TEM data, and P. J. Hirschfeld and Yu. S. Barash for helpful discussion of the Q-reflection. This work is supported by NSF: DMR 07-06013 (HZA), DoE: DE-FG02-07ER46453 (WKP), DoE: DE-AC02-98CH10886 (GDG), and the Center for Emergent Superconductivity, a DoE Energy Frontier Research Center (CRH, LHG, JSW, ZJX).


\end{document}